\documentclass{llncs}
\pdfoutput=1
\usepackage{times}
\usepackage{mathptmx}

\newcommand{\citeauthor}[1]{{AUTHOR!}}
\newcommand{\citet}[1]{{AUTHOR!}~\cite{#1}}
\newcommand{\citep}[1]{\cite{#1}}
\usepackage{verbatim}
\usepackage[caption=false]{subfig}
\usepackage{booktabs}
\usepackage{enumitem}
\usepackage{xspace}
\usepackage{url}
\usepackage[all]{nowidow}
\usepackage{balance}
\usepackage{multirow}
\usepackage{graphicx} 
\def\D{\hphantom{1}}

\newcommand{\myurl}[1]{{\url{#1}}}

\newcommand{\myparagraph}[1]{\vspace{0.40\baselineskip}\noindent{\textbf{#1}}~}
\newcommand\method[1]{{\sf\small{#1}}}

\newcommand{\collection}[1]{\mbox{#1}}

\newcommand{\gov}{\collection{GOV2}}

\newcommand\kb[1]{\mbox{$#1$\,kiB}}
\newcommand\mb[1]{\mbox{$#1$\,MiB}}
\newcommand\gb[1]{\mbox{$#1$\,GiB}}
\newcommand\tb[1]{\mbox{$#1$\,TiB}}

\newcommand{\var}[1]{\mbox{\emph{#1}}}

\newlength{\onedigit}
\newcommand{\offset}{\var{offset}}
\newcommand{\length}{\var{length}}
 
\usepackage[usenames,dvipsnames]{color}

\newcommand{\hoobinpvldb}{Hoobin et al.~\cite{hpz11pvldb}}

\begin{document}

\title{Access Time Tradeoffs in Archive Compression\thanks{Preprint of M. Petri, A. Moffat, P.C. Nagesh, A. Wirth. {\it Access Time Tradeoffs in Archive Compression}, In Proc. Asia Information Retrieval Societies Conference (AIRS) in LNCS vol 9460, pages 15--25, (2015), DOI: 10.1007/978-3-319-28940-3\_2}}
\titlerunning{Tradeoffs in Archive Compression}

\author{
Matthias Petri,
Alistair Moffat,
P. C. Nagesh,
Anthony Wirth
}
\email{}

\institute{Department of Computing and Information Systems\\
The University of Melbourne\\
Victoria 3010, Australia}

\maketitle

\begin{abstract}
Web archives, query and proxy logs, and so on, can all be very large
and highly repetitive; and are accessed only sporadically and
partially, rather than continually and holistically.
This type of data is ideal for compression-based archiving, provided
that random-access to small fragments of the original data can be
achieved without needing to decompress everything.
The recent RLZ (relative Lempel Ziv) compression approach uses a
semi-static model extracted from the text to be compressed, together
with a greedy factorization of the whole text encoded using static
integer codes.
Here we demonstrate more precisely than before the scenarios in which
RLZ excels.
We contrast RLZ with alternatives based on block-based adaptive
methods, including approaches that ``prime'' the encoding for each
block, and measure a range of implementation options using both
hard-disk (HDD) and solid-state disk (SSD) drives.
For HDD, the dominant factor affecting access speed is the
compression rate achieved, even when this involves larger
dictionaries and larger blocks.
When the data is on SSD the same effects are present, but not as
markedly, and more complex trade-offs apply.
\end{abstract}

\section{Introduction}
\label{sec-introduction}

Large data archives are often retained for long periods.
Examples include web crawls; site edit histories for
resources such as the Wikipedia; query, proxy, and click logs; and
many other forms of meta-data associated with the way we store and
access information.
Such archives are rarely decoded in full, and even partial-access
operations may be infrequent.
Moreover, the data might be highly repetitive, with occasional very
long repeated strings, and repeated strings that are widely separated.
There is thus considerable interest in specialized compression
techniques that provide a high level of space saving for such data,
plus the ability to support random access to small fragments of it.

The Relative Lempel-Ziv (RLZ) compression approach is designed for
archives like these~{\citep{hpz11pvldb}}.
It involves a plain-text {\emph{dictionary}} extracted from the
collection of documents via fixed-interval sampling across their
concatenation.
The documents are then factored against the dictionary using the
standard Lempel-Ziv greedy parsing approach, and factor descriptions
consisting of copy offsets and copy lengths are represented with
static integer codes.
Because the dictionary and encodings are both static, decoding is
possible from any point in the encoded stream, provided only that a
corresponding code-aligned byte or bit address is given for the
document that is required.
Moreover, decoding is fast -- during decoding operations the
dictionary is stored in memory uncompressed, allowing rapid access to
factors that can then be copied directly to the output stream as
required.
More details of the RLZ approach are given in Section~\ref{sec-rlz}.

While the approach provided by RLZ is indeed a good solution to the
question of archive compression, other methods based on
{\emph{adaptive}} compression mechanisms are available.
For example, standard tools like {\method{GZip}} and {\method{xz}}
can be applied on a per-block basis.
The block size then becomes an important parameter that trades
compression effectiveness against access speed.
The larger the block size, the better the compression rate, but the
longer it takes for a fragment of text to be reconstructed, since
decompression must start at the beginning of a block.

Our purpose in this paper is to provide detailed evidence of RLZ's
capability in archive compression.
Our analysis includes the effects of the storage device chosen, and
both hard-disk drives (HDD) and solid-state disk (SSD) storage are
employed.
We analyze the factors that determine the time required to access a
fragment of text from an arbitrary location in a large corpus, and
show how different compression techniques can be evaluated.
The approaches explored include making use of a facility provided by
the standard {\method{ZLIB}} library in which a ``priming'' text
enhances compression effectiveness during the start-up phase of
{\method{GZip}}'s Lempel-Ziv implementation.
The various options are compared on the {\gb{426}} {\gov} crawl of
the {\tt{.gov}} domain, which contains a broad mix of HTML, PDF, and
other document formats.

Based on those experiments, we conclude that for HDD the dominant
factor affecting access speed for random decoding is compression
effectiveness, with block size a secondary factor; whereas for SSD
decompression speed is also a factor.
Our results confirm and extend those of {\hoobinpvldb}, providing
additional insights into the behavior of this important archiving
technique.
Our new implementation of RLZ will be made available on completion of
the project, so that other compression approaches can also be
incorporated as they are developed.

\section{RLZ Compression}
\label{sec-rlz}

We now provide a brief description of the RLZ archive compression
mechanism {\citep{hpz11pvldb}}.

\begin{figure}[t]
\centering
\includegraphics[scale=0.8]{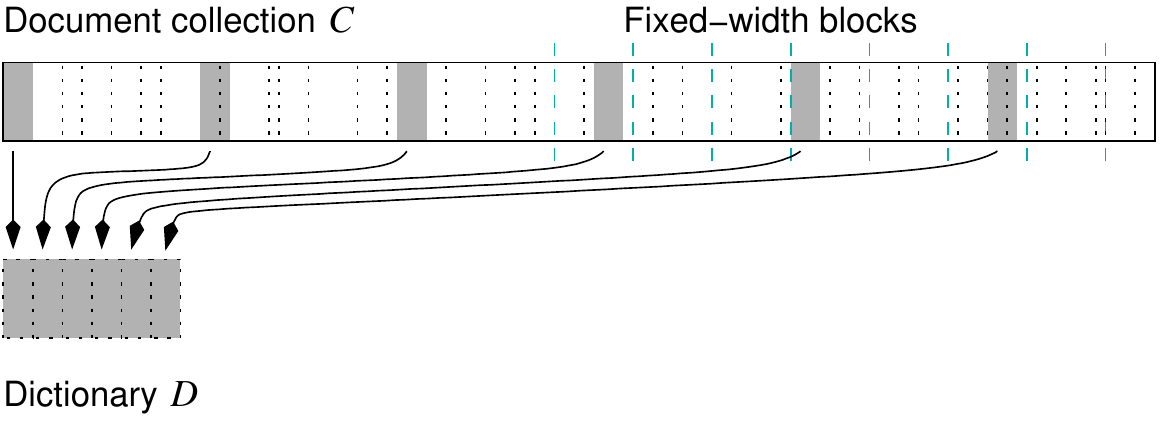}
\caption{Constructing the RLZ dictionary $D$ by selecting regular
samples from the document collection $C$.
Document boundaries in $C$ are shown by dotted lines; block
boundaries (over part of the collection) by dashed lines.
\label{fig-rlz-dictionary}}
\end{figure}

\myparagraph{Forming a Dictionary}
The collection of documents to be stored are concatenated to
make a single large file; we let~$C$ denote that single string,
and~$|C|$ be its length in bytes.
Two parameters are then identified: the {\emph{dictionary size}},
denoted~$|D|$ (with~$D$ to be used for the dictionary); and the
sample size~$s$, chosen to be a factor of~$|D|$.
The dictionary is formed by taking $|D|/s$ samples, each~$s$ bytes
long, from~$C$, extracting them at regular $|C|/(|D|/s)$-byte
intervals.
For example, if $|C|=\gb{64}$ and $s=\kb{1}$, then a dictionary of
$|D|=\mb{64}$ would be formed by concatenating a total of $65{,}536$
samples, extracted every $1{,}048{,}576$ bytes of $C$.
Figure~\ref{fig-rlz-dictionary} shows the process of extracting
regular samples from~$C$ to form the dictionary~$D$, regardless of
the underlying document boundaries.

\myparagraph{Factoring the Collection}
Once~$D$ has been formed, $C$ is broken into a sequence of
{\emph{blocks}}, and each block independently {\emph{factored}}
against~$D$, using a left-to-right greedy approach.
The blocks might be variable-length and formed by considering
individual documents in the collection; might be variable-length and
formed by taking groups of documents to reach some minimum size; or
might be fixed-length and formed by taking some exact number of
bytes.
In our implementation we adopt the latter approach, meaning that
access to any byte range or to any particular document requires that
the corresponding block or blocks be identified and retrieved.

To generate the factorization for each of the blocks, $D$ is indexed
via a suffix array or similar structure, so that for an arbitrary
string~$S$, the set of longest-matching prefixes of~$S$ that appear
in~$D$ can be identified.
Starting at the beginning of each block, factors relative to~$D$ are
identified and represented by a pair of integer values: the
{\emph{length}} of the factor, and its {\emph{offset}} in~$D$.
If the next character in the block does not appear in~$D$, a
{\emph{literal}} is generated -- a factor length of zero, and then an
ASCII character code rather than a dictionary offset.
There are a range of ways in which the presentation of literals can
be optimized, including the application of a minimum match length, or
separating them into a distinct third stream.
These alternatives are explored in Section~\ref{sec-experiments};
{\hoobinpvldb} assume that literals are sufficiently rare that
intermingling them in the stream of offsets will not adversely affect
compression effectiveness.
Except when specifically described otherwise, references to factor
offsets below include any literals that may have been required.
The last factor in each block is truncated so that it finishes at the
block boundary.
The compressed equivalent of each fixed-length block is then the
fundamental access unit for decoding, with higher-level operations
such as document retrieval and byte-range retrieval implemented on
top of the block access routines.

\myparagraph{Compression Rate}
The total cost of storing $C$ is the cost of storing $D$, plus the
cost of storing all of the $\langle\offset,\length\rangle$ pairs.
The dictionary can be stored using any desired compression mechanism,
and is fully decoded into memory prior to subsequent access
operations.
Even stored uncompressed, it is typically a small fraction of the
original collection.
Continuing the previous example, $|D|/|C|=0.1$\%, and a compressed
representation of $D$ should occupy well under $0.03$\% of $|C|$.

The majority of the space required is in the
$\langle\offset,\length\rangle$ pairs.
As already noted, they are separated into two streams on a
per-block basis, with each stream coded using a static method such as
$32$-bit or minimal-width binary integers, or the variable-width
byte-oriented {\method{vbyte}} approach {\cite{wz99compjour}}.
The two coded streams are then typically padded to a byte or word
boundary, concatenated to make a single unit, and a small prelude
added that includes a count of the number of factors contained.
Continuing with the same example, suppose that~$C$ is partitioned
uncompressed into blocks of $\kb{16}$; that the average factor length
is~$20$ bytes; that each $\offset$ is coded in $\log_2|D|=26$ bits;
and that almost all factor lengths are coded in one byte each
({\method{vbyte}} codes for factor lengths of up to~$127$).
Then each factor requires~$34$ bits, and the offsets and lengths for
a block are stored in around $\kb{3.4}$, a compression rate of
approximately~$3.4/16\approx22\%$.
Previous experimental results with RLZ suggest that all these various
estimates are reasonable {\citep{hpz11pvldb}}, and they are further
confirmed in the experiments described in
Section~\ref{sec-experiments}.

\myparagraph{Random Access Decoding}
To provide random-access decoding, index pointers to each block in
the compressed integer stream are maintained in an auxiliary
structure.
The block size determines the number of index points and hence the
size of the index, which is important because the index must also be
retained in memory during access operations.
In the same example, with blocks of $\kb{16}$, a set of
$4{,}194{,}304$ indexing pointers into the compressed stream is
required, with each pointer~$34$ bits long to address a compressed
file of approximately $\gb{16}$.
That is, in the example an index to allow random access to blocks
consumes $\mb{17}$, a further overhead.

To decode a fragment of~$C$ specified by an uncompressed byte range
(for example, if one document is required, and a
mapping from document identifiers to byte addresses is available)
standard mod/div arithmetic is performed to determine the ordinal
numbers of the block or blocks that are required.
The block index (required to be memory resident) is then used to
determine the address of the bundle of de-interleaved
$\langle\offset,\length\rangle$ pairs for that block, and a file
operation undertaken to fetch the relevant data from secondary
storage.
The dictionary~$D$ (also memory resident) is then used, with
$D[\offset]$ to $D[\offset+\length-1]$ copied to a decode buffer for
each $\langle\offset,\length\rangle$ factor extracted from the
compressed blocks.
The required range of bytes from within the block can then be written
to the output stream once the block decode buffer is filled.
That is, after a compressed block has been fetched into main memory,
reconstructing a fragment of~$C$ consists of decoding two sequences
of integers using static integer codes, and then copying strings.
Both operations are fast.
Further blocks are fetched and decoded if required, until the byte
range specified in the query has been delivered.

Ferrada et al.~{\cite{fggp14spire}} have also considered random
access in RLZ mechanisms.

\myparagraph{Memory Footprint}
Compression effectiveness is in part determined by the amount of
space used for the dictionary, as another dimension of
effectiveness-efficiency trade-off.
For example, if the memory required ($\mb{64}+\mb{17}$ in the example
scenario) must be reduced for some reason,
either the block size can be increased, potentially affecting access
speed; or the dictionary size decreased, potentially affecting
compression rate.
If the block size is increased to $\kb{64}$, the index reduces
to~$\mb{4.3}$.
The drawback, of course, is that four times as much data must be
transferred into main memory to fulfill a request, and more of it is
likely to be required to be decoded as well, unless internal
structure is added within each block.
As is demonstrated in the experiments below, transfer and decoding
times are usually small, and block sizes in the tens of kilobyte
range are acceptable.
The uncompressed dictionary~$D$ is then the dominant memory
requirement during random-access decoding.
To mitigate this cost, methods have been developed for pruning the
dictionary to remove unused or under-used
strings~{\citep{twz14sigir}}.

\begin{table}[t]
\setlength{\tabcolsep}{3mm}
\centering
\begin{tabular}{l c c}
\toprule
Medium
		& Random read latency
			& Sequential transfer rate 
\\
\midrule
Hard disk (HDD)
		& $8.5$ milliseconds
			& {\mb{150}}/second
\\
Solid-state disk (SSD)
		& $0.12$ milliseconds
			& {\mb{1000}}/second
\\
					\bottomrule
\\[-0.5ex]
\end{tabular}
\caption{Performance of different storage media.
Extracted from product specifications of current devices: Seagate
ST3000DM001 (HDD), Intel SSD 750 Series (SSD).
\label{tbl-diskstats}}
\end{table}

\myparagraph{Access Time}
In a memory-to-memory context, string-copy decoders similar to RLZ
generate text at around $\mb{250}$--$\mb{300}$ per
second.\footnote{{\small\url{https://github.com/Cyan4973/lz4}}, accessed 27
July 2015.}
A compressed block derived from $\kb{64}$ of~$C$ can thus be decoded
in around~$0.25$ milliseconds.
But that can only happen once it has been fetched from secondary
memory.
Table~\ref{tbl-diskstats} provides indicative performance figures for
mechanical (HDD) and solid-state (SSD) secondary memory devices.
In a mechanical disk, there is a non-trivial startup time for each
data transfer, involving (with high probability) a seek operation to
move the read head, followed by a delay resulting from rotational
latency.
Solid-state disks achieve higher data transfer rates, and commence
the data transfer relatively quickly after the request is received.

If compressed blocks are stored on HDD, the
seek-plus-latency cost of approximately $8.5$~milliseconds
dominates the cost of transferring the data (around~$0.15$
milliseconds for the compressed equivalent of a block of, say,
$\kb{64}$ of~$C$), and the cost of decoding that block once it is in
memory (around $0.25$ milliseconds).
Based on this arithmetic, and assuming that each query consists of
accessing a {\kb{16}} segment of~$C$, a throughput of around $110$
random-access queries per second should be possible.
Of that time, decoding activity occupies less than~$3$\%.
On the other hand, if the whole collection is decoded sequentially
(meaning that seek and latency times are amortized to zero), and if
compression effectiveness of $30$\% or better is achieved (meaning
that decoding cost completely subsumes transfer cost) then data can
be handed to another process at the measured peak output rate.
Continuing the same example, a rate of $\mb{300}$ decoded per second
correspond to up to $5{,}000$ {\kb{64}}-blocks, or $20{,}000$
{\kb{16}}-blocks.

If SSD is used, the situation for random access changes markedly.
Now the transfer initialization time is around~$0.1$ milliseconds,
meaning that something like $2{,}900$ $\kb{64}$ blocks per second can
be fetched and decoded, with the decoding taking around~$60$\% of the
total time.
Sequential access continues to be dominated by decoding cost, and
remains capped at around $20{,}000$ {\kb{16}}-blocks per second.
All of these estimated access time and throughput rates are validated
empirically in Section~\ref{sec-experiments}.

\section{Block-Based Adaptive Alternatives}
\label{sec-gzip}

We now consider additional options for archive compression.

\myparagraph{Standard Compression Libraries}
Standard compression tools such as {\method{GZip}}, {\method{BZip2}},
and {\method{xz}}, are {\emph{adaptive}}, in that they use dynamic
models and codes, so as to be versatile across file types.
For example, the well-known {\method{GZip}} compressor adopts the
same Lempel-Ziv factorization approach as RLZ, starting each
compression run with an empty dictionary, and then adding each parsed
factor's text for possible use in subsequent factorizations.
If {\method{GZip}} is applied independently to blocks, its
``always-start-from-zero'' approach puts it at a disadvantage
compared to RLZ, because the global RLZ dictionary allows
identification of long factors right from the beginning of every
block.

On the other hand, adaptive compression techniques build models that
are focused on exactly the content being compressed, and hence have
an ability to be locally sensitive in a way that RLZ does not.
Adaptive methods are also able to exploit encodings for factor
offsets and lengths that are adaptive rather than static, further
enhancing their ability to provide locally sensitive compression.
That is, while RLZ's use of a global dictionary and static
encodings for factor offsets and lengths gives it an advantage on
very short blocks, localized adaptive methods may obtain better
compression as the block size is increased.
Part of our purpose in this investigation is to explore the options
provided by these alternatives.

\myparagraph{Block Size}
A second area for exploration is the effect of block size.
The connection between block size and the size of the block index was
discussed above.
In the case of RLZ, because it typically uses static integer codes,
increasing block size has no effect on compression effectiveness.
But if large blocks are passed to an adaptive compression utility,
average compression effectiveness is likely to improve, because the
start up cost of the model is amortized over a longer section of
text.
This then raises an interesting trade-off -- at what block size does
an adaptive dictionary provide better compression than a static
RLZ-style dictionary of some given size.

For random-access operations using mechanical disk, the added
decoding cost due a large block size may not matter.
Even with a block size of $\kb{512}$, decoding of half a block, to
reach a given byte address within it, takes around~$0.8$
milliseconds; transfer of a full block takes approximately~$1.1$
milliseconds, assuming a $25$\% compression rate; and the
seek-plus-latency time of around~$8.5$ milliseconds is unchanged.
That is, it should be possible to extract fragments from a block
representing $\kb{512}$ of text in around~$11$ milliseconds, or at an
estimated rate of approximately~$90$ queries per second.

\myparagraph{Batch-Mode Operation}
If queries are batched and processed ``elevator'' style, higher query
throughput rates can be achieved, because average disk-seek times are
likely to be smaller when the access requests are sorted.
For example, if~$110$ random-access queries per second can be
supported without batching, and if batches of sufficient size can be
accumulated so that the average seek-plus-latency time drops from
$8.5$~milliseconds to say $4.5$~milliseconds then the same hardware
configuration should support approximately~$200$ queries per second.
The drawback is that on average the queries will have much greater
latencies before being processed -- perhaps measured in tens or
hundreds of seconds, rather than tens of milliseconds.
In applications that fetch small fragments of a large archive, this
mode of operation may still be acceptable.

\section{Experiments}
\label{sec-experiments}

\myparagraph{A New Implementation}
To allow precise characterization of the performance of RLZ
compression, we have created a new implementation based on
fixed-length data blocks, each compressed independently, with a block
index maintained in memory so that random-access queries can be
supported.
The system is written using $\approx4000$ lines of {\method{C++11}}
code with the help of the {\method{sdsl}}
library~{\cite{gbmp2014sea}}.
We use {\method{gcc 4.9.2}} running on Ubuntu 15.04 in our
experiments, with all optimizations enabled.

We have explored five variants, including three {\method{RLZ}} versions:
\begin{itemize}
\item
	{\method{RLZ-UV}}, using unsigned $32$-bit integers for factor
	offsets, and {\method{vbyte}} for factor lengths, as described by
	{\hoobinpvldb};
\item
	{\method{RLZ-PV}}, using packed $\log_2|D|$-bit integers for
	factor offsets, and {\method{vbyte}} for factor lengths; and
\item
	{\method{RLZ-ZZ}}, using {\method{ZLIB}} (the basis of the standard 
	{\method{GZip}} compression utility) version 1.2.8 (\url{http://zlib.net}) 
	to represent each of the streams of
	$32$-bit factor offsets and the stream of $32$-bit factor
	lengths, on a block-by-block basis.
\end{itemize}
Each of these three methods makes use of a sampled dictionary.
We also applied each of {\method{ZLIB}} and {\method{LZ4}}
(\url{https://github.com/Cyan4973/lz4}) to independent blocks,
without use of a dictionary, following preliminary experimentation
that included {\method{BZip2}} and~{\method{xz}}.
The latter two were slower, and gave less interesting trade-offs
between access speed and compression effectiveness.
Finally, as a sixth system and a further baseline, we measured the
performance of a {\method{COPY}} mechanism that does no compression
at all.

\myparagraph{Datasets}
Our experiments focus on the {\gov} collection, a crawl of the
{\tt{.gov}} domain undertaken in early 2004, with documents stored in
as-crawled order.
This collection contains around~$25$ million documents as a mixture
of PDF, HTML, text, and other formats, averaging $\kb{18}$ each, and
totaling {\gb{426}}.\footnote{
{\small\url{http://ir.dcs.gla.ac.uk/test_collections/gov2-summary.htm}},
27 July 2015.}
We use both the full collection and a {\gb{64}} prefix of it.

\myparagraph{Query Streams}
We explore three modes of retrieval: {\method{FULL}}, in which the
archive is decoded sequentially; {\method{RANDOM}}, in which a
set of $10{,}000$ random unaligned locations is accessed and a
$\kb{16}$ fragment retrieved from each; and {\method{BATCH}}, in
which those same $10{,}000$ locations are accessed, but with the
queries sorted by address.
The ``Sequential'' mode explored by {\hoobinpvldb} most closely
matches our {\method{FULL}} mode, in that they measured retrieval of
$100{,}000$ consecutive {\gov} documents.
Similarly, their ``Query Log'' mode corresponds broadly to our
{\method{RANDOM}} mode, but with $100{,}000$ document requests in the
query stream, and hence more possibility of caching affecting
throughput.

{\hoobinpvldb} also make use of a second URL-sorted {\gov}
collection.
They obtain notably different query throughput results for the two
orderings, particularly with regard to decoding speed, differences
that we were unable to reproduce with our implementation.
An examination of their code suggests that the differences arise from
a mode in their software that because of compiler optimization
inadvertently results in no decoded output being generated.
As a result, we believe that the ``Sequential'' retrieval speeds
shown in their Table~5 (including decoding rates as high as
$80{,}000$ documents per second) should be discounted; and (for other
reasons) possibly some of their other speed results too.\footnote{Our
concerns in this regard have been communicated to the authors of
{\cite{hpz11pvldb}}.}
That is, our work here can be seen in part as representing
re-measurement of the techniques {\hoobinpvldb} describe.

\myparagraph{Dictionary Size and Formation}
The effectiveness of the RLZ mechanism is heavily affected by the
dictionary size.
In their {\gov} experiments {\hoobinpvldb} work with dictionary sizes
between {\gb{0.5}} and {\gb{2}}.
Here we focus on smaller dictionaries, and explore the range from
$\mb{16}$ to $\mb{256}$ for the $\gb{64}$ test file, and the range
$\mb{64}$ to $\mb{1024}$ for the full {\gov} collection.
As described in Section~\ref{sec-rlz}, we followed the ``standard''
approach of selecting fixed-interval samples from the collection,
presuming it to have been concatenated into a single large file.
Other dictionary construction methodologies have been shown to result
in small compression effectiveness gains {\citep{twz14sigir}}; we
also explored a range of other heuristics, but found the simple
interval-based sampling approach to be relatively robust.
We used samples of length $s=1024$ throughout, matching (when
$|D|=\gb{1}$) some of the experiments carried out by {\hoobinpvldb}.
We tested block sizes of $\kb{16}$, $\kb{64}$, and $\kb{256}$.
All compression rates include the cost of storing the dictionary,
compressed as a character stream using {\method{ZLIB}}, and the cost
of the index table for block access, also stored using
{\method{ZLIB}}.

\myparagraph{Hardware Configuration}
All experiments were run on a server equipped with two Intel Xeon
E5640 CPUs running at $2.67$~GHz using {\mb{144}} RAM, a Western
Digital {\tb{5}} (WD50EFRX-68MYMN1) HDD and a {\gb{500}} Samsung 850
EVO SSD.
Before each experiment the operating system caches were cleared to
minimize caching effects using {\tt{echo~3~>~/proc/sys/vm/drop\_caches}}.
We also took care with file placement on the HDD, noting the effect
that fragmentation and track assignment can have on disk-based
experimentation {\cite{wm05adcs}}.
In some cases this meant deleting and re-copying indexes, so as to
ensure that measurements were made in a fair and consistent manner.
The SSD did not suffer from this variability.

\begin{figure}[t]
\centering
\hspace*{-0.5em}\includegraphics[width=1.02\textwidth]{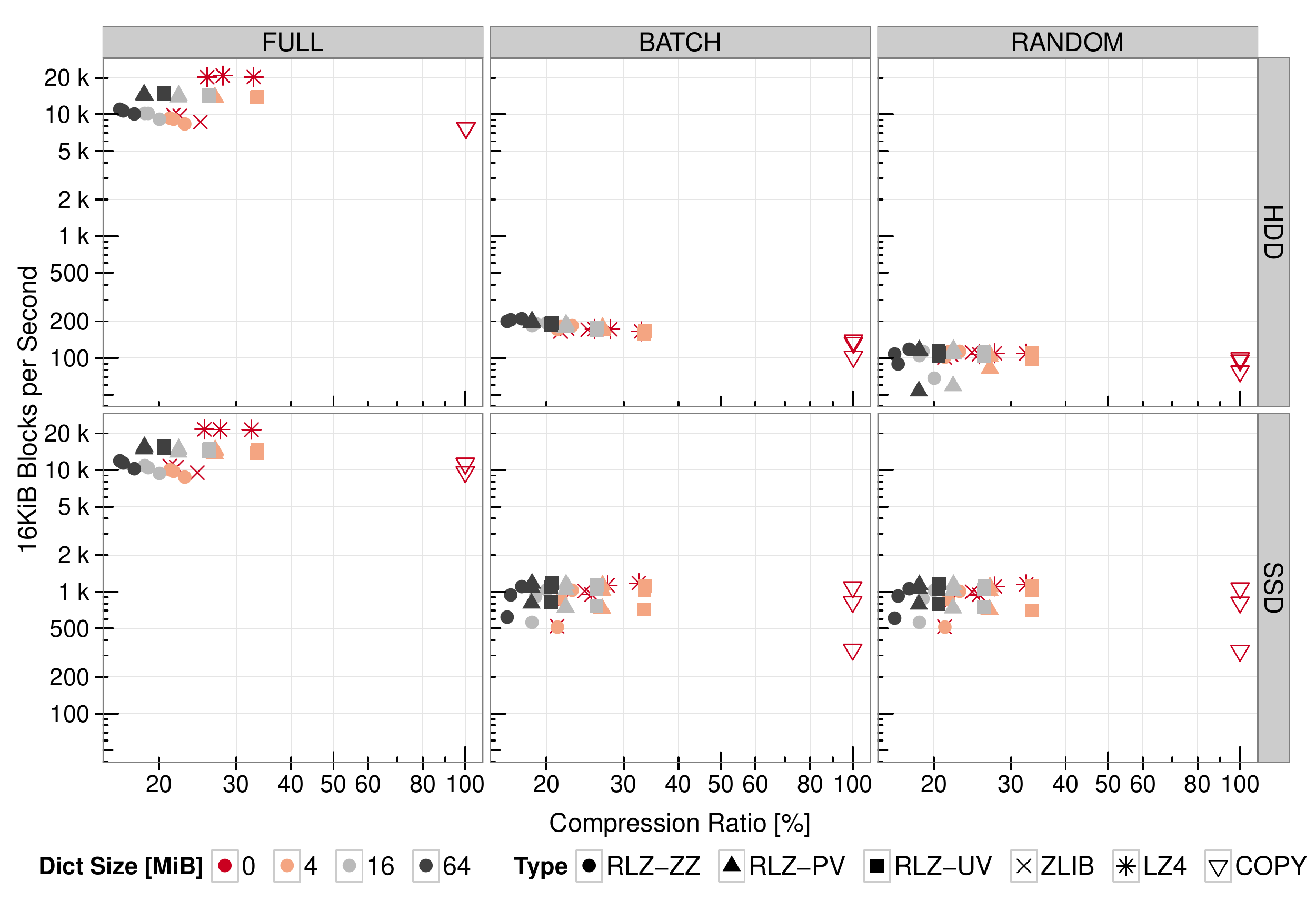}
\caption{Query processing rates measured as {\kb{16}} units retrieved
per second, for three different processing modes, two types of
secondary storage, block sizes of {\kb{16}}, {\kb{64}}, and
{\kb{256}} (not individually identified in the plots), and a
$\gb{64}$ prefix of {\gov}.
In the {\method{FULL}} mode, throughput rates are for aligned
{\kb{16}} units; for the {\method{BATCH}} and {\method{RANDOM}}
modes, for unaligned {\kb{16}} units.
The {\method{COPY}}, {\method{LZ4}}, and {\method{ZLIB}} methods do
not use a dictionary, and are shown as {\mb{0}}.
In general, larger block sizes lead to better compression
effectiveness; together with faster access in the case of
{\method{FULL}} operation, and slower access in the case of
{\method{BATCH}} and {\method{RANDOM}} operation.
\label{fig-rlz-overview}}
\end{figure}

\myparagraph{High-Level View}
Figure~\ref{fig-rlz-overview} presents an overview of the six
methods, measured using the $\gb{64}$ file, and shows the gross
relative performance across the three querying modes and two hardware
configurations.
Each pane plots the relationship between compression rate, as a
percentage of the original file size, on the horizontal axis; and
access speed, measured by the number of $\kb{16}$ blocks accessed per
second.
Each pane contains~$36$ plotted points: three RLZ variants, each with
three different dictionary sizes and three different block sizes
($27$ data points); plus two blocked adaptive methods using the same
three different block sizes ($6$ data points); plus the
{\method{COPY}} method using the three block sizes.
Each color corresponds to a dictionary size, and each point shape
corresponds to a method.
Within each method, the larger the dictionary size and/or the larger
the block size, the better the compression.
But increased block sizes also correspond to slower decoding.
All six panes show the absolute advantage of using virtually any
compression method, with the {\method{COPY}} approach the slowest
in several cases, and never the fastest.
Data compression often pays for itself.
Note also that for each method, dictionary, and block size combination the
compression rate is the same across all six panes.

The two left panes confirm that sequential decoding is very fast,
with the {\method{LZ4}}, {\method{RLZ-UV}} and {\method{RLZ-PV}}
approaches having a moderate speed advantage over the other
mechanisms, but with all of the compressed approaches delivering
$10{,}000$+ documents (each a $\kb{16}$ unit in these experiments)
per second, or $\mb{160}$+/second.
There is little measurable difference in performance between HDD and
SSD.
Unsurprisingly, the larger the dictionary and/or the larger the block
size, the better the compression.

The {\method{BATCH}} and {\method{RANDOM}} modes are much slower.
In the two middle panes, depicting {\method{BATCH}} access, there is
a clear trend on the HDD for better compression to correspond to
higher query throughput, with query rates of between $100$ documents
(unaligned $\kb{16}$ units in this querying mode) and $200$
documents per second, and relatively little differentiation between
the compression techniques.
On the SSD, much faster rates of $800$--$2{,}000$ documents per
second result, with throughput more sensitive to the
choice of compression technique.
Finally, the right two panes show the further slowdown arising from
{\method{RANDOM}} access.
On the HDD, query rates are around $100$ documents/second; and
on the SSD querying throughput is the same as for {\method{BATCH}}
retrieval.

The SSD {\method{RANDOM}} and {\method{BATCH}} querying rates are
around half those predicted by the model described in
Section~\ref{sec-rlz}.
Measurement of the operating characteristics of the SSD used in the
experiments indicate that its mean latency is higher than is shown in
Table~\ref{tbl-diskstats}, approximately $0.25$ millseconds per
access, explaining the difference between predicted and measured
querying rates.

\begin{figure}[t]
\centering
\hspace*{-0.5em}\includegraphics[width=1.02\textwidth]{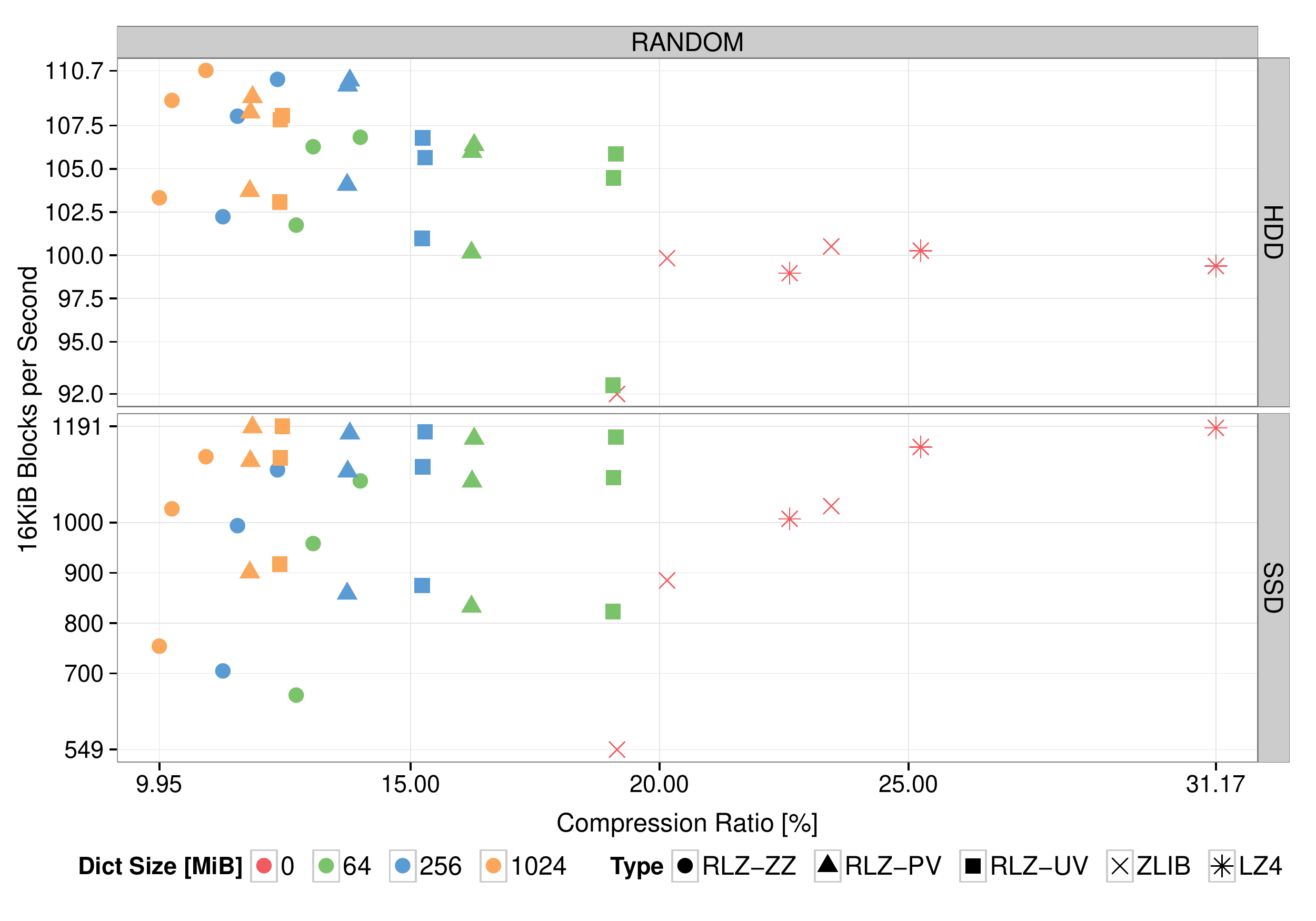}
\caption{Query processing rates for the {\method{RANDOM}} processing
mode, measured as unaligned $\kb{16}$ units retrieved per second, for
two types of secondary storage, block sizes of {\kb{16}}, {\kb{64}},
and {\kb{256}} (not individually identified in the plots), and the
full {\gov} collection.
Note that the upper and lower panes have different vertical scales.
\label{fig-rlz-details-rand}}
\end{figure}

\myparagraph{Detailed View -- Random Access}
Figure~\ref{fig-rlz-details-rand} shows a focused view corresponding
to the two right-hand panes in Figure~\ref{fig-rlz-overview},
measured using the full {\gb{426}} {\gov} collection, and with the
{\method{COPY}} method omitted.
It considers only the {\method{RANDOM}} queries, using correspondingly
larger dictionaries of $\mb{64}$, $\mb{256}$, and $\gb{1}$, and
unchanged block sizes of $\kb{16}$, $\kb{64}$, and $\kb{256}$.
At the increased scale of these graphs, it is possible to identify a
Pareto frontier for each different dictionary size, and quantify the
tension between compression and throughput that is controlled by
block size.

For random access, the raw speed of {\method{LZ4}} is less of an
advantage, and it is part of the trade-off frontier only when no
dictionary can be used, and when the fast data rates of SSD are
available.
If dictionary space is not a restriction, then the {\method{RLZ-ZZ}}
methods dominate absolutely for HDD retrieval, and for much of the
frontier with SSD retrieval.
The remaining part of the SSD frontier is pinned on the
{\method{RLZ-PV}} method, highlighting that unaligned bit-wise
integers can be processed just as efficiently as can the aligned
$32$-bit integers preferred by {\hoobinpvldb}, and give better
compression.

Comparing our results with those of {\hoobinpvldb}, we have measured
very similar throughput rates for {\method{RANDOM}} queries, and by
adding blocking to the {\method{RLZ-ZZ}} approach, have slightly
improved its compression effectiveness.
That small gain, and the reduction in transfer and decoding time that
accompanies it, gives the {\method{RLZ-ZZ}} approaches the upper
hand, and dictionaries as small as {\mb{256}} are sufficient to
attain high {\method{RANDOM}} query throughput even compared to
{\method{RLZ-PV}}, and also compact storage.
On SSD, the situation is similar, but if query throughput is the
primary goal, the {\method{RLZ-PV}} represent the best combination of
attributes.

\section{RLZ Extensions}

We briefly describe two different ways in which {\method{RLZ}}
compression can be enhanced.

\begin{table}[t]
\renewcommand{\tabcolsep}{0.4em}
\centering
\begin{tabular}{c c cc c cc c cc c cc}
\toprule
Block
	&& \multicolumn{2}{c}{\method{ZLIB}} 
		&& \multicolumn{2}{c}{\method{ZLIB}$'$}
			&& \multicolumn{2}{c}{\method{RLZ-ZZ}} 
				&& \multicolumn{2}{c}{\method{RLZ-ZZ}$'$}
\\
\cmidrule{3-4}\cmidrule{6-7}\cmidrule{9-10}\cmidrule{12-13}
size
	&& comp. & thrpt.
		&& comp. & thrpt.
			&& comp. & thrpt.
				&& comp. & thrpt.
\\
\midrule
{\kb{\D16}}
	&& 24.83\% &\D990
		&& 22.64\% &\D955
			&& 17.56\% & 1043
				&& 17.37\% &\D946
\\
{\kb{\D64}}
	&& 22.29\% &\D840
		&& 21.53\% &\D825
			&& 16.56\% &\D905
				&& 16.47\% &\D866
\\
{\kb{256}}
	&& 21.53\% &\D513
		&& 21.33\% &\D508
			&& 16.26\% &\D599
				&& 16.21\% &\D581
\\
\bottomrule
\vspace*{-1.0ex}
\end{tabular}
 
\caption{Use of {\method{ZLIB}} priming with the {\gb{64}} prefix of
{\gov}.
In the {\method{ZLIB}}$'$ method, a uniform sampled dictionary of
{\mb{256}} is employed.
In the {\method{RLZ-ZZ}}$'$ method, the same {\mb{256}} dictionary is
used, plus two fixed pre-computed integer sequences of {\kb{64}}
containing factor lengths and factor offsets respectively.
The two values for each combination are the compression rate, as a
percentage of the original collection, and the measured
{\method{RANDOM}}-mode throughput, in documents per second using SSD.
\label{tbl-priming}}
\end{table}

\myparagraph{Priming in RLZ-ZZ}
The {\method{ZLIB}} compression library offers the ability to
``prime'' the compression process, by providing data that is
considered to precede the sequence that is to be compressed, thereby
providing a model to initialize the dictionary.
In the same way that {\method{RLZ}} employs a dictionary, so too can
a {\method{ZLIB}$'$} approach, in which a uniform sampled dictionary
is created, and then each block of data is {\method{ZLIB}}-compressed
using priming text drawn from the dictionary in the vicinity of
the block being compressed.
A similar approach has been demonstrated to be effective when
compressing Yahoo email archives {\citep{bz15dcc}}.
A primed variant of {\method{RLZ-ZZ}} can also be constructed, using
pre-computed sequences of factor offsets and factor lengths.
Table~\ref{tbl-priming} shows that when the block size is small, priming
achieves a worthwhile benefit, but that the gain for larger block
sizes is smaller.
Priming causes a small decrease in query throughput rates.

\myparagraph{Three Streams}
Using a full factor -- requiring $30$+ bits -- to represent a literal
is expensive, and it is not actually necessary for literals to be
mingled with the stream of dictionary offsets.
If a third stream is added, containing only the sequence of literals,
it can be compressed separately.
Once a separate stream is allowed, it also makes sense to force any
short factors in to it too -- if the next match in the dictionary is
of length less than some value $\var{min\_literal}$, then the entire
factor is coded as literals.
Similar optimizations are used in many Lempel-Ziv implementations;
see, for example, Fiala and Greene~\cite{fg89cacm}.
The third stream can be coded using any of the mechanisms already
discussed, or any other coding method {\cite{mt02caca}}; here we use
of {\method{ZLIB}} for all three.

\begin{table}[t]
\renewcommand{\tabcolsep}{0.4em}
\centering
\begin{tabular}{c c ccc c cccc}
\toprule
Block
	&& \multicolumn{3}{c}{\method{RLZ-ZZ}} 
		&& \multicolumn{4}{c}{\method{RLZ-ZZZ}} 
\\
\cmidrule{3-5}\cmidrule{7-10}
size
	&& {\mb{16}} & {\mb{64}} & {\mb{256}}
		&& {\mb{16}} & {\mb{64}} & {\mb{256}} & thrpt.
\\
\midrule
{\kb{\D16}}
	&& 22.89\% & 20.03\%& 17.56\%
		&& 22.42\% & 19.80\%& 17.47\% & 1029
\\
{\kb{\D64}}
	&& 21.58\% & 18.89\%& 16.57\%
		&& 20.99\% & 18.54\%& 16.39\% &\D896
\\
{\kb{256}}
	&& 21.18\% & 18.54\%& 16.27\%
		&& 20.57\% & 18.17\%& 16.06\% &\D591
\\
\bottomrule
\vspace*{-1.0ex}
\end{tabular}

\caption{Use of a three-way split of streams, using
$\var{min\_literal}=4$, a $\gb{64}$ prefix of {\gov}, and three
different dictionary sizes.
Values reported are compression rates, as a percentage of the
original collection.
The final column shows the measured {\method{RANDOM}}-mode
throughput, as unaligned $\kb{16}$ accesses per second using SSD
secondary storage, for the {\method{RLZ-ZZZ}} method with a
dictionary of {\mb{256}}, and can be compared with the values in
Table~\ref{tbl-priming}.
\label{tbl-threestreams}}
\end{table}

Table~\ref{tbl-threestreams} provides a detailed comparison between
{\method{RLZ-ZZ}} and {\method{RLZ-ZZZ}}.
The gain in compression is larger with a small dictionary than with a
large dictionary, since the bigger the dictionary, the less likely it
is that short factors will get generated.
That is, the use of three streams can be viewed as being a way of
making slightly better use of a small dictionary.
Decoding speed is only marginally affected.

\section{Summary and Conclusion}
\label{sec-conclusion}

We have extended the experimentation of {\hoobinpvldb} to SSD memory,
and undertaken a systematic study of blocking effects and access time
trade offs in archive compression.
The {\method{RLZ-ZZ}} static-dictionary method provides an
outstanding balance between random access query throughput and
compression effectiveness, for both HDD devices and SSD devices.
We have also measured the effect of two simple techniques that
provide small additional compression gains, without any great loss of
throughput.

\myparagraph{Acknowledgments}
This work was supported under the Australian Research Council's
{\emph{Discovery Projects}} scheme (project DP140103256).
We have had access to the code of Hoobin~et~al.\ while working on
this project, and we thank them for making it available.

\vspace{-0.25cm}
\begin{small}
\bibliographystyle{splncs03}
\bibliography{strings-shrt,local}

\begin{thebibliography}{1}
\providecommand{\url}[1]{\texttt{#1}}
\providecommand{\urlprefix}{URL }

\bibitem{bz15dcc}
Bergman, A., Zohar, E.: Compressing {Y}ahoo mail. In: Proc. DCC. pp. 223--232
  (2015)

\bibitem{fggp14spire}
Ferrada, H., Gagie, T., Gog, S., Puglisi, S.J.: Relative {L}empel-{Z}iv with
  constant-time random access. In: Proc. SPIRE, pp. 13--17 (2014)

\bibitem{fg89cacm}
Fiala, E.R., Greene, D.H.: Data compression with finite windows. Comm. ACM
  32(4),  490--505 (1989)

\bibitem{gbmp2014sea}
Gog, S., Beller, T., Moffat, A., Petri, M.: From theory to practice: Plug and
  play with succinct data structures. In: Proc. SEA. pp. 326--337 (2014)

\bibitem{hpz11pvldb}
Hoobin, C., Puglisi, S.J., Zobel, J.: Relative {L}empel-{Z}iv factorization for
  efficient storage and retrieval of web collections. PVLDB  5(3),  265--273
  (2011)

\bibitem{mt02caca}
Moffat, A., Turpin, A.: Compression and Coding Algorithms. Kluwer, Boston
  (2002)

\bibitem{twz14sigir}
Tong, J., Wirth, A., Zobel, J.: Principled dictionary pruning for low-memory
  corpus compression. In: Proc. SIGIR. pp. 283--292 (2014)

\bibitem{wm05adcs}
Webber, W., Moffat, A.: In search of reliable retrieval experiments. In: Proc.
  10th Australasian Document Computing Symp. pp. 26--33 (2005)

\bibitem{wz99compjour}
Williams, H.E., Zobel, J.: Compressing integers for fast file access. Comp. J.
  42(3),  193--201 (1999)

\end{thebibliography}
\end{small}

\end{document}